\documentclass{llncs}

\usepackage{amsmath}
\usepackage{amssymb}
\usepackage{graphicx}
\usepackage{bm}
\usepackage{enumerate}
\usepackage{float}
\usepackage{tikz}
\usetikzlibrary{chains}
\usepackage{pgfplots}
\usepackage{algorithm}
\usepackage{algorithmicx}
\usepackage[noend]{algpseudocode}

\usepackage{makecell}
\usepackage{booktabs}
\usepackage{multirow}
\usepackage{url}
\newtheorem{thm}{Theorem}
\newtheorem{defi}{Definition}
\def\bx{\bm{x}}
\def\bw{\bm{w}}
\def\be{\bm{e}}
\def\bc{\bm{c}}
\def\by{\bm{y}}
\def\bs{\bm{s}}
\def\bt{\bm{t}}
\def\C{\mathcal{C}}
\def\N{\mathbb{N}}
\def\F{\mathbb{F}}

\begin{document}
\title{Syndrome decoding meets multiple instances}

\author{Haoxuan Wu\inst{1,2} and  Jincheng Zhuang\inst{1,2}}
\institute{
Key Laboratory of Cryptologic Technology and Information Security,\\
Ministry of Education, Shandong University, Qingdao 266237, China \\
\and
School of Cyber Science and Technology, Shandong University,\\  Qingdao 266237, China \\
}
\maketitle
\pagestyle{plain}

\begin{abstract}
The NP-hard problem of decoding random linear codes is crucial to both coding theory and cryptography. In particular, this problem underpins the security of many code based
post-quantum cryptographic schemes. The state-of-art algorithms for solving this problem
are the information syndrome decoding algorithm and its advanced variants. In this work, we consider syndrome decoding in the multiple instances setting. Two strategies are applied for different scenarios. The first strategy is to solve all instances with the aid of the precomputation technique.
We adjust the current framework and distinguish the offline phase and online phase to reduce the amortized complexity.
Further, we discuss the impact on the concrete security of some post-quantum schemes.
The second strategy is to solve one out of many instances.
Adapting the analysis for some earlier algorithm,
we discuss the effectiveness of using advanced variants and confirm a related folklore conjecture.

\end{abstract}

\section{Introduction}

\noindent\emph{Coding theory.}
A binary linear code $\C$ is a $k-$dimensional subspace of $\F_2^{n}$.
In the decoding problem, the receiver obtains the word $\bx = \bc+\be$, where $\bc$ is a codeword and $\be$ is the error vector with Hamming weight $\omega$.
The goal is to find $\be$ and thus recovering the codeword $\bc$.
Generally, the weight $\omega$ is bounded by the minimum distance $d$ of the code $\C$ for Full Distance Decoding and the distance $d/2$ for Half Distance Decoding.

The state-of-art algorithm for decoding random linear codes is a type of algorithm called Information Set Decoding(ISD) and its advanced variants. The ISD algorithm was first proposed by Prange~\cite{Pra62} in 1962. The time complexity for the worst case is $2^{0.121n}$. Then this algorithm is enhanced by Stern~\cite{stern88} and Dumer~\cite{Dum91} to $2^{0.117n}$. Later, May, Meurer and Thomae(MMT)~\cite{MMT11}~improved previous results to $2^{0.112n}$. In 2012, Becker, Joux, May and Meurer(BJMM)~\cite{BJMM12} reduced the time complexity in the worst case to $2^{0.102n}$ by using some special representation technique.

Using a new algorithm that solves the nearest neighbor search problem, May and Ozerov(MO)~\cite{MO15}~modified the ISD algorithm in 2015 to accelerate the ISD decoding method.
The BJMM algorithm~\cite{BJMM12} mentioned above with this new technique was accelerated to $2^{0.0967n}$ and was subsequently revised in~\cite{BM17b}~as $2^{0.0953n}$.  Furthermore, Both and May(BM)~\cite{BM18} presented an algorithm utilizing tree structure, combing the nearest neighbor search in each step. Their work is currently asymptotically the best algorithm for decoding random linear codes.

In 2022, Esser and Bellini(EB)~\cite{EB22}~proposed a general framework that includes essentially all the above advanced ISD decoding algorithms, and revisited them from the nearest neighbor perspective.
Additionally, they provided an estimator ~\cite{SyndromeDecodingEstimator} that makes it easy to estimate the concrete time complexity.

On July 5th 2022, NIST~\cite{NIST}~announced the candidates to be standardized and round 4 submissions in the post-quantum schemes standardization process.
There are three code-based schemes~\cite{McEliece,Bike,HQC}~in the fourth round, and their difficulty is related to the decoding problem of random linear codes.
During the NIST standardization process, it is essential to determine the concrete security parameters of each scheme, which depends on the algorithms of decoding.

\noindent\emph{Preprocessing technique.}
Algorithms utilizing the preprocessing idea consist of two stages.
During the preprocessing phase (also known as off-line phase), adversary can access public information, which he uses to calculate auxiliary information for later phase.
Then comes the online phase.
Adversary is provided with concrete instances and he aims to apply the previously information to solve these instances.

Preprocessing technique has been applied for solving different problems which underlying the difficulty of public-key encryption scheme, like~\cite{mihalcik10,BL13,C-GK18,CDG18}~in the discrete logarithm problem and the algorithm~\cite{Laa16,DLW19,DLvW20}~for solving closest vector problem (CVP).
For the discrete logarithm problem, Mihalcik~\cite{mihalcik10}, Bernstein and Lange~\cite{BL13} proposed a general preprocessing method.
Following that, Corrigan-Gibbs and Kogan~\cite{C-GK18}~proved lower bound estimates
under certain restrictions.
Laarhoven~\cite{Laa16} proposed an algorithm for solving of CVP
with the preprocessing technique. Subsequently, the above algorithm is improved by~\cite{DLW19,DLvW20}.

The preprocessing technology is especially beneficial in the multiple instances setting, since the algorithm is divided into preprocessing steps and online processing steps.
In this way, it is possible to reduce the amortization complexity of solving each instance by sharing the same preprocessing information.

\noindent\emph{Decoding One Out of Many~(DOOM).}
The idea of the scalability of public key encryption was recently considered by Auerbach, Giacon and Kiltz~\cite{AGK20}.
They took into account the connection between the time complexity of solving one instance and the time complexity of solving $M$ instances when given $N$ instances. $M = 1$ corresponds to
the one out of many scenario.
In the one out of many instances setting for decoding problem, the adversary succeeds when one out of $N$ independent decoding instances is solved, where the instances have the same parity check matrix.
In 2009, Finiasz and Sendrier~\cite{FS09} analyzed this problem and proposed an algorithm for syndrome decoding using a generalized birthday algorithm.
In 2011, Sendrier~\cite{sendrier11} designed an algorithm to solve the above problem using the ISD algorithm and speed up Stern's algorithm by a factor of roughly $\sqrt{N}$.
As remarked in~\cite{EB22}, a folklore conjecture concerning decoding one out of many is that such acceleration applies for general variants, but no work has been conducted to examine decoding one out of many for advanced ISD decoding such as~\cite{BJMM12,BM18} in detail.

\subsection{Our work}
This work focuses on the syndrome decoding problem in the multiple instance setting.
There are two main contributions of this work.
\begin{itemize}
\item First, we apply the preprocessing technique to ISD that divides the decoding process into a preprocessing and an online process for solving syndrome decoding problems.
After that, we analyze the complexity of each phase.
As a result, the amortization complexity of solving the multiple-instance problem is reduced.
Furthermore, we consider the impact to the concrete security of code based post quantum schemes.
\item Second, we consider resolving one out of $N$ instances.
We examine advanced ISD  decoding variants by adjusting the work of Sendeier~\cite{sendrier11}.
A theoretical analysis for the time complexity estimation of some specific schemes confirm the folklore conjecture.
The time complexity of solving one out of $N$ instances is roughly as $1/\sqrt{N}$ of the time complexity of solving a single instance. This has applications in estimating the message complexity of some code-based schemes as in~\cite{EB22}.
\end{itemize}

This paper is organized as follows.
In Section 2, we introduce the necessary notations and review some decoding algorithms.
In Section 3, we propose a decoding algorithm framework with preprocessing adapted from Esser-Bellini's work~\cite{EB22}.
Then, we apply this framework to estimate the concrete parameter security of some post quantum schemes.
In section 4, we describe the algorithm in one out of many setting for advanced ISD algorithms and analyze the complexity.
In the last section, we conclude the paper.

\section{Preliminaries}

Denote matrices by capital letters and column vectors by lowercase bold letters. Let $I_n$ be the $n\times n$ identity matrix.
We define $[d] := \{1,2,\cdots,d\}$ for an integer $d\in\N$. For a vector $\bw\in\F_2^n$ we denote the Hamming weight of $\bw$ as $wt(\bw) := |\{i\in[n]|w_i = 1\}|$.
Moreover, denote $\mathcal{B}_p^n := \{\bm{w}\in\F_2^n|wt(\bm{w}) = p\}$ as the set of all vectors of length $n$ and weight $p$.

A binary linear code $\C$ of length $n$ and dimension $k$ is a $k-$dimensional space of $\F_2^n$. It can be defined by a parity check matrix $H\in\F_2^{(n-k)\times n}$ via
\[
\C := \{\bc\in\F_2^n|H\bc = 0\}.
\]
The minimum distance of linear code $\C$ is defined by $d:=min_{\bc\neq \bc'\in\C}{\Delta(\bc,\bc')}$, where $\Delta(\bc,\bc')$ is the number of distinct coordinates between $\bc$ and $\bc'$. A random binary code means that each element in the parity check matrix is chosen uniformly at random from $\F_2$.

For a faulty word $\by\in\F_2^n$, it can be written as $\by = \bc+\be$ for $\bc\in\C$ and an error $\be$. Then
\begin{equation}\label{equ:synd}
H\by = H(\bc+\be) = H\be = \bs.
\end{equation}
The vector $\bs$ is called the \emph{syndrome} of $\by$.

\begin{defi}
  \textbf{(syndrome decoding problem).} Given the parity check matrix $H$ of a random binary linear code $\C$, $\bs\in\F_2^{n-k}$ and $\omega\in[d]$. The syndrome decoding problem is finding a vector $\be\in\F_2^n$ satisfying $H\be = \bs$ with $wt(\be) = \omega$.
\end{defi}

The state-of-art algorithms that we know today for decoding random binary linear codes are so-called \emph{Information Set Decoding}(ISD) algorithms. This class of algorithms is based on the algorithm introduced by Prange in 1962~\cite{Pra62}~.

Prange's algorithm first performs a random column permutation of the parity check matrix $H$. For an arbitrary permutation $P$, the equation $H\be = \bs$ can be transformed into $(HP)(P^{-1}\be) = \bs$. Then we apply the Gaussian elimination to the matrix $HP$ to obtain the systematic form. More precisely, this is done by multiplying by an invertible matrix $Q\in\F_2^{(n-k)\times(n-k)}$
\begin{equation*}
Q(HP) = (\tilde{H}I_{n-k})\text{, for }\tilde{H}\in\F_2^{(n-k)\times k}.
\end{equation*}
Furthermore, assume that $(P^{-1}\be) = (\be',\be'')\in\F_2^k\times\F_2^{n-k}$ and $Q\bs = \tilde{\bs}$, then Equation~(\ref{equ:synd}) is transformed as
\begin{equation*}
Q(HP)(P^{-1}\be) = \tilde{H}\be'+\be'' = Q\bs = \tilde{\bs}.
\end{equation*}

Suppose that after permutation, $wt(\be') = p$, and hence $wt(\be'') = \omega-p$. Then the decoding problem is transformed into finding a vector $\be'\in\F_2^k$ with $wt(\be') = p$ satisfying
\[
wt(\be'') = wt(\tilde{H}\be'+\tilde{\bs}) = \omega-p,
\]
which means that
\begin{equation*}
  \tilde{H}\be' = \tilde{\bs} \text{ on all }n-k \text{ but }\omega-p \text{ coordinates}.
\end{equation*}

Once we find the vector $\be'$ that matches the condition, we can solve the syndrome decoding problem by computing $\be = P(\be',H\be'+\tilde{\bs})$.
In Prange's algorithm, the parameter $p = 0$, while the parameter $p$ is set to be a positive number in all subsequent improvements.

In Dumer's ISD algorithm~\cite{Dum91}~, a new parameter $l$ was introduced and the parity check matrix $H$ was transformed into a new form
\[
Q(HP) = \begin{pmatrix}
          \tilde{H}_1 & \bm{0} \\
          \tilde{H}_2 & I_{n-k-l}
        \end{pmatrix}\text{ , where } \tilde{H}_1\in\F_2^{l\times(k+l)}\text{ and } \tilde{H}_2\in\F_2^{(n-k-l)\times(k+l)}.
\]
Let $\tilde{\bs} = Q\bs = (\bs_1,\bs_2)\in\F_2^{l}\times\F_2^{n-k-l}$ and $P^{-1}\be = (\be',\be'')\in\F_2^{k+l}\times\F_2^{n-k-l}$. Then Equation~(\ref{equ:synd}) can split as
\begin{equation}\label{equ:ISD-Dum1}
  \tilde{H}_1\be' = \bs_1,
\end{equation}
\begin{equation}\label{equ:ISD-Dum2}
  \tilde{H}_2\be' + \be'' = \bs_2.
\end{equation}

Note that Equation~(\ref{equ:ISD-Dum1})~is an \emph{exact} matching problem and Equation~(\ref{equ:ISD-Dum2})~ is an \emph{approximate} matching problem. Assume the weight distribution satisfies $wt(\be') = p$ and $wt(\be'') = \omega-p$. The syndrome decoding problem is transformed  as follows. First, find $\be'$ with $wt(\be') = p$ that satisfies Equation~(\ref{equ:ISD-Dum1}). Then, check whether the solution meets $wt(\tilde{H}_2\be'+\bs_2) = \omega-p$, and if the equality holds, the solution is $P(\be',\tilde{H}_2\be'+\bs_2)$.

Dumer used a meet-in-the-middle strategy to solve the exact matching problem of Equation~(\ref{equ:ISD-Dum1}).
The BJMM algorithm~\cite{BJMM12}~improves the previous result of Dumer~\cite{Dum91}.
In a nutshell, they split Equation~(\ref{equ:ISD-Dum1})~into $\tilde{H}_1\be'_{1} + \tilde{H}_1\be'_{2}=\bs_1$, then determine the solution of this equation by a three-level binary search tree.
The result $\be'_{1},\be'_{2}$ is then checked by Equation~(\ref{equ:ISD-Dum2}).

For an approximate matching problem, May and Ozerov~\cite{MO15}~presented a nearest neighbor search problem and the corresponding algorithm.
Given two vector lists $L_1,L_2$, the nearest neighbor search problem aims to find all $(\bx_1,\bx_2)\in (L_1,L_2)$ such that $wt(\bx_1+\bx_2)$ equals to some specified value.
For the decoding problem, the verification of $\be'_{1},\be'_{2}$ via Equation~(\ref{equ:ISD-Dum2})~can be accelerated by the May-Ozerov nearest neighbor search algorithm.

In 2018, Both and May~\cite{BM18}~presented a new algorithm for the decoding of random binary linear codes.
They achieved the current optimal asymptotic complexity by combing the ISD and the nearest neighbor techniques introduced in ~\cite{MO15}.

Esser and Bellini~\cite{EB22} proposed a general information set decoding framework in 2022 that uses the nearest neighbor search technique, incorporating essentially all previous ISD algorithms. Additionally, they provide estimator tools to analyze the actual overhead of various ISD algorithms, providing estimates of the security parameters of the coding-based NIST candidates.

Next, we briefly review the Esser-Bellini's framework. Assume that the parity matrix has a systematic form, and it can be written as follows
\begin{equation}\label{mat:H-pre}
H = (\tilde{H},I_{n-k}) = \begin{pmatrix}
                            \tilde{H_1} & I_{l_1} & \bm{0} & \bm{0} \\
                            \tilde{H_2} & \bm{0} & I_{l_2} & \bm{0} \\
                            \tilde{H_3} & \bm{0} & \bm{0} & I_{l_3}
                          \end{pmatrix}\text{, where }\tilde{H}\in\F_2^{(n-k)\times k}.
\end{equation}
Then split the syndrome $\tilde{\bs}$ (after Gaussian elimination) into
\begin{equation}\label{equ:s}
\tilde{\bs} = (\tilde{\bs}_1,\tilde{\bs}_2,\tilde{\bs}_3)\in\F_2^{l_1}\times\F_2^{l_2}\times\F_2^{l_3}.
\end{equation}
The distribution of the weights of the solution $\be$ changes after the permutation. It can be expressed as follows: $\be = (\be',\be_{1}'',\be_{2}'',\be_{3}'')\in(\F_2^{k}\times\F_2^{l_1}\times\F_2^{l_2}\times\F_2^{l_3})$ and
\begin{equation}\label{equ:e}
wt(\be') = p \text{ and } wt(\be_{i}'') = \omega_{i} \text{ for }i=1,2,3.
\end{equation}

The above parameters can be numerically analyzed to obtain optimal values that satisfy the relationship $l_3 = n-k-l_1-l_2$ and $\omega_3 = \omega-p-\omega_1-\omega_2$.
Thus the original syndrome decoding problem's equation can be rewritten as
\begin{equation}\label{equ:pre}
\tilde{H}_i\be' = \tilde{\bs}_i \text{ on all }l_i \text{ but } \omega_i \text{ coordinates for }i=1,2,3.
\end{equation}

As a result, the syndrome decoding problem is reduced to three approximate matching problems.
Essentially, they split $\be' = \be_1 + \be_2$ with $\be_1,\be_2 \in\F_2^k$ and $wt(\be_{1(\text{resp. }2)}) = p_1$ for some optimal value concerning $p$.
Firstly, find the set of vectors satisfying Equation~(\ref{equ:pre}) for $i=1$,
and then sieve the vectors in this set using the $i=2$ conditions to obtain a smaller set.
In the end, verification of the obtained set can be done by considering the original equation in the syndrome decoding.
Additionally, they use a clever tree structure to find the vectors that match the conditions.

Preprocessing was also a part of the previous ISD decoding technique.
It is feasible, for instance, to compute the permutation matrix and Gaussian elimination steps in advance.
However, our proposal pre-processes more data.

\section{ISD Algorithm from a preprocessing perspective}

The framework of Esser-Bellini~\cite{EB22}~described above is able to simulate previously introduced ISD algorithms such as BJMM~\cite{BJMM12}, May-Ozerov~\cite{MO15}, and Both-May~\cite{BM18} by changing the specific implementation of their steps. In this section, we combine this algorithmic framework~\cite{EB22}~with the preprocessing technique, and consider its impact on the concrete parameter security of some code based post-quantum schemes.

Next, we analyze the framework for binary search tree in depth 2 in detail from a preprocessing point of view. To begin with, construct four sets $L_{i}, i = 1,2,3,4$.
\begin{equation}\label{equ:layer2}
\begin{aligned}
    &L_i = \{(\bx_i||\tilde{H}\bx_i) \in\F_2^{k/2}\times0^{k/2}\times\F_2^{n-k}| wt(\bx_i) = p_1/2\text{ for }i=1,3\},\\
    &L_i = \{(\bx_i||\tilde{H}\bx_i) \in 0^{k/2}\times\F_2^{k/2}\times\F_2^{n-k}| wt(\bx_i) = p_1/2\text{ for }i=2,4\}.
\end{aligned}
\end{equation}

These four sets are the candidate sets that satisfy Equation~(\ref{equ:pre}) for $i=1$, and we need to find the vectors that satisfy the other conditions simultaneously.
The process is as follows.
For $\tilde{H}\bx_{1(2)}$ part of the set $L_{1(2)}$, we apply the nearest neighbor search technique to find vector pairs that satisfy $wt(\tilde{H}\bx_{1}+\tilde{H}\bx_{2}) = \omega_{11}$ in the first $l_1$ coordinates and include them in the set $L_{12}$. The set $L_{12}$ has the form
\begin{equation*}
    L_{12} = \{\bx_1 + \bx_2||\tilde{H}(\bx_1 + \bx_2)|wt(\tilde{H}(\bx_1 + \bx_2)) = \omega_{11} \text{ in the first } l_1 \text{ coordinates}\}.
\end{equation*}

Each of the steps outlined above can be completed in the \emph{preprocessing step}.
More specifically, when the parity check matrix and the weight of the solution are known, the rest of the parameters can be determined relatively,
which means that both the set $L_i,i=1,2,3,4$ and
the set $L_{12}$ can be calculated in advance for one time.

When we know the target vector $\bs$ for the required solution,
we can proceed to the \emph{online computation step},
where operations are related to the target vector.
First, to find vectors that satisfy $wt(\tilde{H}\bx_{1}+\tilde{H}\bx_{2})+\tilde{\bs}_1 = \omega_{11}$,
we apply the nearest neighbor search technique to the $H\bx_i,i=3,4$ part in sets $L_3$ and $L_4$  and then construct the set $L_{34}$. The set $L_{34}$ has the form
\begin{equation*}
  L_{34} = \{\bx_3 + \bx_4||\tilde{H}(\bx_3 + \bx_4)|wt(\tilde{H}(\bx_3 + \bx_4)+\bs) = \omega_{11} \text{ in the first } l_1 \text{ coordinates}\}.
\end{equation*}

We are now presented with two sets of candidate vectors $L_{12}$ and $L_{34}$, which are collections of all candidate $\be_1$ and $\be_2$, i.e., $\bx_1+\bx_2$ and $\bx_3+\bx_4$.
The list $L_{1234}$ is then obtained by using the nearest neighbor search technique again, which is the set of candidate vectors of $\be'$.

\subsection{ISD algorithm with preprocessing}

\begin{algorithm}
\caption{ISD-NN-Preprocessing Framework(Adapted from~\cite{EB22})}\label{alg:frame}
\begin{algorithmic}[1]
    \Require parity check matrix $H\in\F_2^{(n-k)\times n}$, syndrome $\bs$, error weight $\omega\in[n]$
    \Ensure $\be\in\F_2^{n}$, which satisfies $H\be = \bs$ and $wt(\be) = \omega$

    Compute parameters: $l_1,l_2,\omega_1,\omega_2,\omega_{11},p_1,p$
    \State Parameters are based on the definition in Equation~(\ref{mat:H-pre}),(\ref{equ:s}),(\ref{equ:e})
    \Repeat
    \Repeat\quad let index $j= 0$ \Comment{\emph{Preprocessing Phase}}
        \State $j++$
        \State Choose random permutation matrix $P^{(j)}\in\F_2^{n\times n}$
        \State $HP^{(j)}$ is transformed into systematic form by using Gaussian elimination with a invertible matrix $Q^{(j)}$. Obtain $\tilde{H}_{1}^{(j)},\tilde{H}_{2}^{(j)}$ and $\tilde{H}_{3}^{(j)}$
        \State Construct 4 lists $L_1^{(j)},L_2^{(j)},L_3^{(j)}$ and $L_4^{(j)}$ by the definition~\ref{equ:layer2}, precisely
        \[
        L_i^{(j)} = \{\bx_i|\bx_i = (\bm{y},0^{k/2}):\bm{y}\in\mathcal{B}_{p_1/2}^{k/2}\} \text{ for }i=1,3
        \]
        \[
        L_i^{(j)} = \{\bx_i|\bx_i = (0^{k/2},\bm{y}):\bm{y}\in\mathcal{B}_{p_1/2}^{k/2}\} \text{ for }i=2,4
        \]
        \State Compute $L_{12}^{(j)}$ by $L_1^{(j)}$ and $L_2^{(j)}$, $L_{12}^{(j)}$ has the form
        \[
        L_{12}^{(j)} = \{\bx_1 + \bx_2||\tilde{H}^{(j)}(\bx_1 + \bx_2)|(\bx_1,\bx_2)\in L_1\times L_2\text{ and }wt(\tilde{H}_1^{(j)}\bx_{1}+\tilde{H}_1^{(j)}\bx_{2}) = \omega_{11}\}
        \]
    \Until{$j = T_P$}
    \Repeat\quad let index $j=0$ \Comment{\emph{Online Phase}}
        \State $j++$
        \State Compute $L_{34}^{(j)}$ by $L_3^{(j)}$ and $L_4^{(j)}$, $L_{34}^{(j)}$ has the form
        \[
        L_{34}^{(j)} = \{\bx_3 + \bx_4||\tilde{H}^{(j)}(\bx_3 + \bx_4)|(\bx_3,\bx_4)\in L_3\times L_4\text{ and }wt(\tilde{H}_1^{(j)}\bx_{3}+\tilde{H}_1^{(j)}\bx_{4} + \tilde{\bs}_1) = \omega_{11}\}
        \]
        \State Compute $L_{1234}^{(j)}$ by $L_{12}^{(j)}$ and $L_{34}^{(j)}$, $L_{1234}^{(j)}$ has the form
        \[
        L_{1234}^{(j)} = \{\bx_{12} + \bx_{34}||\tilde{H}^{(j)}(\bx_{12} + \bx_{34})|(\bx_{12},\bx_{34})\in L_{12}^{(j)}\times L_{34}^{(j)}\text{ and }wt(\tilde{H}_2^{(j)}\bx_{3}+\tilde{H}_2^{(j)}\bx_{4} + \tilde{\bs}_2) = \omega_{2}\}
        \]
        \For{$\bx\in L_{1234}$}
                $\tilde{\be} = (\bx,\tilde{H}\bx + \tilde{\bs})$
            \If{$wt(\be) = \omega$}
                Break the \emph{online phase}
            \EndIf
        \EndFor
    \Until{$j=T_P$}
    \Until{$wt(\be) = \omega$}
    \State\Return The corresponding $P^{(j)}\tilde{\be}$
\end{algorithmic}
\end{algorithm}

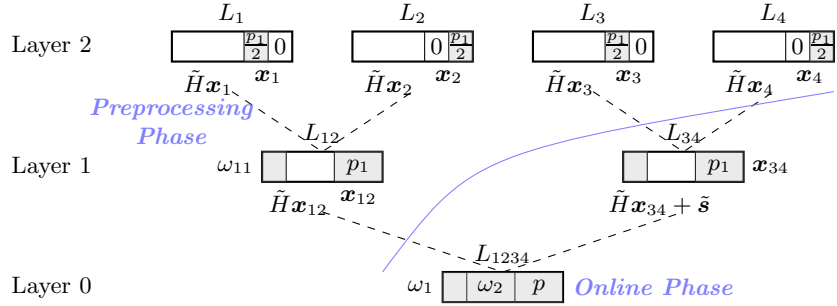
\begin{figure}
\centering
\begin{tikzpicture}[scale = 0.8]
    \draw[thick,-](0,0)--(2,0)--(2,-0.5)--(0,-0.5)--cycle;
    \draw[thick,-](3,0)--(5,0)--(5,-0.5)--(3,-0.5)--cycle;
    \draw[thick,-](6,0)--(8,0)--(8,-0.5)--(6,-0.5)--cycle;
    \draw[thick,-](9,0)--(11,0)--(11,-0.5)--(9,-0.5)--cycle;
    \draw[-](1.2,0)--(1.2,-0.5);
    \draw[-](1.6,0)--(1.6,-0.5);
    \draw[-](4.2,0)--(4.2,-0.5);
    \draw[-](4.6,0)--(4.6,-0.5);
    \draw[-](7.2,0)--(7.2,-0.5);
    \draw[-](7.6,0)--(7.6,-0.5);
    \draw[-](10.2,0)--(10.2,-0.5);
    \draw[-](10.6,0)--(10.6,-0.5);
    \filldraw[fill= gray!50,opacity=0.3] (1.2,0) -- (1.6,0) -- (1.6,-0.5) -- (1.2,-0.5) -- cycle;
    \filldraw[fill= gray!50,opacity=0.3] (4.6,0) -- (5,0) -- (5,-0.5) -- (4.6,-0.5) -- cycle;
    \filldraw[fill= gray!50,opacity=0.3] (7.2,0) -- (7.6,0) -- (7.6,-0.5) -- (7.2,-0.5) -- cycle;
    \filldraw[fill= gray!50,opacity=0.3] (10.6,0) -- (11,0) -- (11,-0.5) -- (10.6,-0.5) -- cycle;

    \node at (-2,-0.25){Layer 2};
    \node at (1.4,-0.25){$\frac{p_1}{2}$};
    \node at (4.8,-0.25){$\frac{p_1}{2}$};
    \node at (7.4,-0.25){$\frac{p_1}{2}$};
    \node at (10.8,-0.25){$\frac{p_1}{2}$};
    \node at (1.8,-0.25){$0$};
    \node at (4.4,-0.25){$0$};
    \node at (7.8,-0.25){$0$};
    \node at (10.4,-0.25){$0$};
    \node[above] at (1,0){$L_1$};
    \node[above] at (4,0){$L_2$};
    \node[above] at (7,0){$L_3$};
    \node[above] at (10,0){$L_4$};
    \node[below] at (0.6,-0.5){$\tilde{H}\bx_1$};
    \node[below] at (1.6,-0.5){$\bx_1$};
    \node[below] at (3.6,-0.5){$\tilde{H}\bx_2$};
    \node[below] at (4.6,-0.5){$\bx_2$};
    \node[below] at (6.6,-0.5){$\tilde{H}\bx_3$};
    \node[below] at (7.6,-0.5){$\bx_3$};
    \node[below] at (9.6,-0.5){$\tilde{H}\bx_4$};
    \node[below] at (10.6,-0.5){$\bx_4$};

    \draw[thick,-](1.5,-2)--(3.5,-2)--(3.5,-2.5)--(1.5,-2.5)--cycle;
    \draw[thick,-](7.5,-2)--(9.5,-2)--(9.5,-2.5)--(7.5,-2.5)--cycle;
    \draw[-](2.7,-2)--(2.7,-2.5);
    \draw[-](8.7,-2)--(8.7,-2.5);
    \draw[-](1.9,-2)--(1.9,-2.5);
    \draw[-](7.9,-2)--(7.9,-2.5);
    \filldraw[fill= gray!50,opacity=0.3] (1.5,-2) -- (1.9,-2) -- (1.9,-2.5) -- (1.5,-2.5) -- cycle;
    \filldraw[fill= gray!50,opacity=0.3] (2.7,-2) -- (3.5,-2) -- (3.5,-2.5) -- (2.7,-2.5) -- cycle;
    \filldraw[fill= gray!50,opacity=0.3] (7.5,-2) -- (7.9,-2) -- (7.9,-2.5) -- (7.5,-2.5) -- cycle;
    \filldraw[fill= gray!50,opacity=0.3] (8.7,-2) -- (9.5,-2) -- (9.5,-2.5) -- (8.7,-2.5) -- cycle;

    \node at (-2,-2.25){Layer 1};
    \node at (3.1,-2.25){$p_1$};
    \node at (9.1,-2.25){$p_1$};
    \node[left] at (1.5,-2.25){$\omega_{11}$};
    \node[below] at (2.1,-2.5){$\tilde{H}\bx_{12}$};
    \node[below] at (8.1,-2.5){$\tilde{H}\bx_{34}+\tilde{\bs}$};
    \node[below] at (3.1,-2.5){$\bx_{12}$};
    \node[right] at (9.5,-2.25){$\bx_{34}$};
    \node[above] at (2.5,-2){$L_{12}$};
    \node[above] at (8.5,-2){$L_{34}$};
    \draw[dashed] (1,-1)--(2.5,-2)--(4,-1);
    \draw[dashed] (7,-1)--(8.5,-2)--(10,-1);

    \draw[thick,-](4.5,-4)--(6.5,-4)--(6.5,-4.5)--(4.5,-4.5)--cycle;
    \draw[-] (4.9,-4)--(4.9,-4.5);
    \draw[-] (5.7,-4)--(5.7,-4.5);
    \filldraw[fill= gray!50,opacity=0.3] (4.5,-4)--(6.5,-4)--(6.5,-4.5)--(4.5,-4.5)--cycle;
    \node at (6.1,-4.25){$p$};
    \node at (5.3,-4.25){$\omega_2$};
    \node[left] at (4.5,-4.25){$\omega_1$};
    \node[above] at (5.5,-4){$L_{1234}$};
    \draw[dashed] (2.5,-3)--(5.5,-4)--(8.5,-3);
    \node at (-2,-4.25){Layer 0};

    \draw[color = blue!50] (11,-1) .. controls (5,-2) .. (3.5,-4);
    \node[align = center,color = blue!50] at(0,-1.5){\textbf{\textit{Preprocessing}}\\\textbf{\textit{Phase}}};
    \node[color = blue!50] at(8,-4.25){\textbf{\textit{Online Phase}}};
\end{tikzpicture}
\caption{Illustration of ISD algorithm with preprocessing}\label{fig:frame}
\end{figure}

We assume $T_P$ steps are required to obtain the weight distribution of solution $\be$ in Equation~(\ref{equ:e}).
The Algorithm~\ref{alg:frame} is the ISD decoding framework which uses preprocessing idea adapted from~\cite{EB22}.
Pictorially, Figure~\ref{fig:frame} illustrates the ISD process with preprocessing intuitively.

\subsection{Analysis of the preprocessing framework}
Frist, the correctness analysis of our algorithm follows from that of~\cite{EB22}.
We refer readers for more details in the Section 4 of ~\cite{EB22}.
Next, we analyze the time complexity of the framework, which is similar to that of ~\cite{EB22}. But here we distinguish the  preprocessing phase and the online phase.
Denote $R$ as the number of representations of $\be' = \be_1 + \be_2$ with $wt(\be_i) = p_1$.
Define $q$ as the probability that the above representation satisfies the $L_{12}$ construction.
Let $T_G$ represent the time of Gaussian elimination for each initial parity check matrix $H$.
For the list with $L$ elements, the time for nearest neighbor search with the length of $L$ and weight of $\omega$ is denote by $\mathcal{N}_{L,l,\omega}$.

In the preprocessing phase, we compute $T_P$ permutations as required.
For each permutation, we need to use the Gaussian elimination to obtain the specified parity check.
Compared to building the $L_{12}$ set, the base sets $L_i,i=1,2,3,4$ construction takes less time.
So the time complexity for preprocessing phase is
\[
T_{pre} = T_P\cdot\left(T_G+\max(1,(R\cdot q)^{-1})\cdot\mathcal{N}_{|L_1|,l_1,\omega_{11}}\right)
\]

During the online computation phase, we only need to construct the sets $L_{34}$ and $L_{1234}$, so the time complexity for the online phase is
\[
T_{Online} = T_P\cdot\max(1,(R\cdot q)^{-1})\cdot\left(\mathcal{N}_{|L_1|,l_1,\omega_{11}}+\mathcal{N}_{|L_{12}|,l_2,\omega_{2}}\right)
\]

The complexity analysis method above can be used to analyze all specific algorithms based on this framework.

\subsection{Concrete parameters}
Various ISD decoding algorithms can be obtained by instantiating the framework above with specific parameters.
In this section, based on Esser and Bellini's open source estimator~\cite{SyndromeDecodingEstimator}, we evaluate the bit security of code-based schemes in the 4th round of NIST PQC standardization process.
NIST PQC standardization process defines five levels of security parameters, corresponding to the difficulty of solving AES128, AES192, and AES256, respectively.
Table 1 of~\cite{EB22} provides specific security parameters for code based cryptographic schemes at different levels.
The purpose of this section is to analyze the bit security of three different schemes using preprocessing algorithms.
For each algorithm, we analyze the total complexity $T$ and the online phase complexity $T_{online}$ using \emph{Both-May}, \emph{May-Ozerov}, \emph{BJMM}, \emph{BJMM-P-DW}, and \emph{BJMM-DW} respectively.

\begin{table}
  \centering
  \renewcommand\arraystretch{1.5}
  \caption{Bit security for Classic McEliece scheme.}\label{tab:McE}
  \begin{tabular}{ccccccccccc}
    & \multicolumn{2}{c}{\makecell[c]{Category 1\\($n$ = 3488)}} & \multicolumn{2}{c}{\makecell[c]{Category 3\\($n$ = 4608)}} & \multicolumn{2}{c}{\makecell[c]{Category 5\\($n$ = 6688)}} & \multicolumn{2}{c}{\makecell[c]{Category 5\\($n$ = 6960)}} & \multicolumn{2}{c}{\makecell[c]{Category 5\\($n$ = 8192)}} \\
    \cline{2-11}
     & $T$ & $T_{online}$ & $T$ & $T_{online}$ & $T$ & $T_{online}$ & $T$ & $T_{online}$ & $T$ & $T_{online}$ \\
    \midrule[1pt]
    \emph{Both-May} & 143 & 142 & 182 & 182 & 250 & 249 & 249 & 249 & 281 & 281 \\
    \emph{May-Ozerov} & 141 & 141 & 180 & 180 & 246 & 246 & 246 & 245 & 276 & 275 \\
    \emph{BJMM} & 142 & 142 & 183 & 182 & 248 & 248 & 248 & 247 & 278 & 277 \\
    \emph{BJMM-P-DW} & 143 & 143 & 183 & 182 & 249 & 248 & 248 & 248 & 279 & 277 \\
    \emph{BJMM-DW} & 144 & 143 & 183 & 182 & 250 & 250 & 250 & 249 & 282 & 279 \\
    \hline
  \end{tabular}
\end{table}


In Table~\ref{tab:McE}, the bit security is provided for the suggested parameters of the classical McEliece scheme.
Tables~\ref{tab:BIKE} and~\ref{tab:HQC} show the estimates for the BIKE and HQC schemes, respectively.



\begin{table}[h]
  \centering
  \renewcommand\arraystretch{1.5}
  \caption{Bit security for BIKE scheme.}\label{tab:BIKE}
  \begin{tabular}{cccccccc}
    \multicolumn{2}{c}{} & \multicolumn{2}{c}{\makecell[c]{Category 1\\($n$ = 24646)}} & \multicolumn{2}{c}{\makecell[c]{Category 3\\($n$ = 49318)}} & \multicolumn{2}{c}{\makecell[c]{Category 5\\($n$ = 81946)}} \\
    \cline{3-8}
    \multicolumn{2}{c}{} & $T$ & $T_{online}$ & $T$ & $T_{online}$ & $T$ & $T_{online}$ \\
    \midrule[1pt]
    \multirow{5}*{message} & \emph{Both-May} & 147 & 144 & 212 & 210 & 276 & 276 \\
     & \emph{May-Ozerov} & 146 & 143 & 211 & 209 & 276 & 274  \\
     & \emph{BJMM} & 147 & 143 & 211 & 209 & 277 & 274 \\
     & \emph{BJMM-P-DW} & 147 & 143 & 211 & 209 & 276 & 274 \\
     & \emph{BJMM-DW} & 147 & 145 & 211 & 210 & 277 & 276 \\
    \hline
    \multirow{5}*{key} & \emph{Both-May} & 148 & 146 & 211 & 210 & 278 & 278 \\
     & \emph{May-Ozerov} & 147 & 145 & 210 & 209 & 278 & 277 \\
     & \emph{BJMM} & 147 & 145 & 211 & 209 & 279 & 277 \\
     & \emph{BJMM-P-DW} & 147 & 145 & 211 & 209 & 278 & 277 \\
     & \emph{BJMM-DW} & 147 & 146 & 211 & 210 & 279 & 278 \\
    \hline
  \end{tabular}
\end{table}

By virtue of the cyclic structure of the BIKE scheme, Esser and Bellini distinguish message security and key security in their estimator.
The two kinds of security can benefit from this structure in different ways.
In Table~\ref{tab:BIKE}, we also distinguish between the message security and the key security.


\begin{table}
  \centering
  \renewcommand\arraystretch{1.5}
  \caption{Bit security for HQC scheme.}\label{tab:HQC}
  \begin{tabular}{ccccccc}
     & \multicolumn{2}{c}{\makecell[c]{Category 1\\($n$ = 35338)}} & \multicolumn{2}{c}{\makecell[c]{Category 3\\($n$ = 71702)}} & \multicolumn{2}{c}{\makecell[c]{Category 5\\($n$ = 115274)}} \\
    \cline{2-7}
     & $T$ & $T_{online}$ & $T$ & $T_{online}$ & $T$ & $T_{online}$ \\
    \midrule[1pt]
    \emph{Both-May} & 146 & 142 & 214 & 211 & 276 & 273 \\
    \emph{May-Ozerov} & 145 & 141 & 214 & 210 & 276 & 272 \\
    \emph{BJMM} & 146 & 141 & 214 & 210 & 276 & 272 \\
    \emph{BJMM-P-DW} & 146 & 141 & 214 & 210 & 276 & 272 \\
    \emph{BJMM-DW} & 146 & 144 & 214 & 213 & 276 & 275 \\
    \hline
  \end{tabular}
\end{table}

\begin{remark}
In the table of complexity analysis, the values for time complexity are all logarithms with a base of 2, making the distinction between $T$ and $T_{online}$ less noticeable.
The differences between the two will be more noticeable if we utilize a deeper tree structure, such as depth 3 or  depth 4 tree structure, and optimize respective parameters.
An intuitive explanation is that the most time-consuming step, the nearest neighbor search, occurs three times in the 2-level tree structure. We can only pre-compute one of them.
As a contrast, four out of  seven of the nearest neighbor searches that occur in the three-layer decoding process can be pre-computed.
As an illustration, this is confirmed by experimental data shown in Table~\ref{tab:depth} using~\cite{SyndromeDecodingEstimator}.

\end{remark}

\begin{table}
  \centering
  \renewcommand\arraystretch{1.5}
  \caption{Online complexity with different depth for Classic McEliece scheme.}\label{tab:depth}
  \begin{tabular}{ccccccccccc}
    & \multicolumn{2}{c}{\makecell[c]{Category 1\\($n$ = 3488)}} & \multicolumn{2}{c}{\makecell[c]{Category 3\\($n$ = 4608)}} & \multicolumn{2}{c}{\makecell[c]{Category 5\\($n$ = 6688)}} \\
    \cline{2-7}
     & \makecell[c]{$T_{online}$\\(depth = 2)} & \makecell[c]{$T_{online}$\\(depth = 3)} & \makecell[c]{$T_{online}$\\(depth = 2)} & \makecell[c]{$T_{online}$\\(depth = 3)} & \makecell[c]{$T_{online}$\\(depth = 2)} & \makecell[c]{$T_{online}$\\(depth = 3)} \\
    \midrule[1pt]
    \emph{BJMM} & 142 & 141 & 182 & 180 & 248 & 247 \\
    \emph{May-Ozerov} & 141 & 140 & 180 & 180 & 246 & 246 \\
    \hline
  \end{tabular}
\end{table}

\section{Decoding one out of many for advanced ISD variants}
In the decoding one out of many~(DOOM) problem, one is given $N$ syndrome decoding instances from the same code, and aims to solve one of them. This problem stems from certain scenario such as
in public key cryptography, where the receiver can obtain multiple instances for the same linear code, and solve one of them is meaningful.

In 2009, Finiasz and Sendrier~\cite{FS09} analyzed this problem and proposed an algorithm for syndrome decoding using a generalized birthday algorithm. In 2011, Sendrier~\cite{sendrier11} designed an algorithm to solve the problem using the ISD algorithm and speed up Stern's algorithm by a factor of roughly $\sqrt{N}$.

Briefly, Sendrier first presented an abstract framework for the previous ISD decoding algorithm (like Stern's algorithm~\cite{stern88}) in which two sets $W_1$ and $W_2$ are constructed.
A solution to the syndrome decoding problem can be obtained by comparing the elements in the two sets.
Based on this abstraction framework, Sendrier proposed increasing the size of set $W_1$ while decreasing the size of set $W_2$ when $N$ instances are given.
Compared to solving one instance, the time complexity can be decreased by $\sqrt{N}$ when solving one of the $N$ objectives
for certain variant like Stern's algorithm~\cite{stern88}.

As remarked in~\cite{EB22}, for more advanced ISD decoding algorithms, such as BJMM~\cite{BJMM12}, there is currently no analytical result comparable to the above. Nevertheless,
a folklore conjecture is that similar speed up applies.
And such result is utilized in the estimator of Esser-Bellini.
In this section, we fill this analytical gap, and provide theoretical support for the conjecture.

\subsection{Generalized framework of advanced ISD variants}
Combining the idea of~\cite{sendrier11} and of~\cite{EB22},
we first establish the generalized framework of the advanced ISD variants.
Specifically, we replace the 1 layer search in~\cite{sendrier11} to decode one of many instances by a modified multi-layer search proposed in~\cite{EB22}.
The size of the base sets $L_{i},i=1,2,3,4$ is the same in for~\cite{EB22}, and may be different in the current framework.
The generalized information set decoding framework is shown in Algorithm~\ref{alg:gene-frame}.
The notations used here follow the same as Section 2.
In particular, the four sets $L_i\subset \mathcal{B}_{p_i}^{k}$ for $i = 1,2,3,4$ and $p_1+p_2+p_3+p_4 = p$, where $|L_1||L_2||L_3||L_4| = \binom{k}{p}$.

\begin{algorithm}[h]
\caption{Generalized ISD Framework (Adapted from~\cite{sendrier11,EB22})}\label{alg:gene-frame}
\begin{algorithmic}[1]
    \Require parity check matrix $H\in\F_2^{(n-k)\times n}$, syndrome $\bs$, error weight $\omega\in[n]$
    \Ensure $\be\in\F_2^{n}$, which satisfies $H\be = \bs$ and $wt(\be) = \omega$
    \Repeat
    \State choose random permutation matrix $P\in\F_2^{n\times n}$
    \State apply Gaussian elimination to transform $HP$ and $\bs$ into Equation~(\ref{mat:H-pre})
    \ForAll{$e_1\in L_1$ and $e_2\in L_2$}
    {
    \\
    \quad\quad\quad compute set $L_{12}$ which has form
    \[
    L_{12} \leftarrow \{\be_1 + \be_2|(\be_1,\be_2)\in L_1\times L_2\text{ and }wt(\tilde{H}_1\be_{1}+\tilde{H}_1\be_{2}) = \omega_{11}\}
    \]
    }
    \EndFor
    \ForAll{$e_3\in L_3$ and $e_4\in L_4$}
    {
    \\
    \quad\quad\quad compute set $L_{34}$ which has form
    \[
    L_{34} \leftarrow \{\be_3 + \be_4|(\be_3,\be_4)\in L_3\times L_4\text{ and }wt(\tilde{H}_1\be_{3}+\tilde{H}_1\be_{4})+ \tilde{\bs_1} = \omega_{11}\}
    \]
    }
    \EndFor
    \ForAll{$\bx_1\in L_{12}$ and $\bx_2\in L_{34}$}
    {
    \\
    \quad\quad\quad compute set $L_{1234}$ which has form
    \begin{equation*}
    L_{1234} \leftarrow \{\bx_1+\bx_2|(\bx_1,\bx_2)\in L_{12}\times L_{34}\text{ and }wt(\tilde{H}_2\bx_{1}+\tilde{H}_2\bx_{2})+ \tilde{\bs_2} = \omega_{2}\}
    \end{equation*}
    }
    \EndFor
    \For{$\bx\in L_{1234}$}
    \\
    \quad\quad\quad compute $(\bx,\tilde{H}\bx+\tilde{\bs})$
    \If{$wt(\bx,\tilde{H}\bx+\tilde{\bs}) = \omega$}
    \\
    \quad\quad\quad\quad\quad Break
    \EndIf
    \EndFor
    \Until{$wt(\bx,\tilde{H}\bx+\tilde{\bs}) = \omega$}
    \State The corresponding $P(\bx,\tilde{H}\bx+\tilde{\bs})$
\end{algorithmic}
\end{algorithm}

Assume that $P$ random permutations are needed to obtain the weight distribution, $T_G$ is the time for the Gaussian elimination, and $T_{12}, T_{34}$, and $T_{1234}$ is the time required to construct the sets $L_{12}, L_{34}$ and $L_{1234}$ respectively.
Denote $R$ as the number of different representations into which $\be$ might be split, and $q$ as the probability that such a representation meets the restrictions of the sets $L_{12}$ and $L_{34}$.

The total time complexity is as follows
\begin{equation}\label{equ:time-DOOM}
  T = P\times{T_G} + P\times\max\left(1,(R\cdot q)^{-1}\right)\left(T_{12}+T_{34}+T_{1234}\right).
\end{equation}

Let $\varepsilon$ be the probability that $\be'\in\mathcal{B}_p^k$ is the part of the final solution $\be = (\be',\be''_1,\be''_2,\be''_3)$. Then

\begin{equation*}
  \varepsilon \approx \frac{\binom{l_1}{\omega_1}\binom{l_2}{\omega_2}\binom{n-k-l_1-l_2}{\omega-p-\omega_1-\omega_2}}{\binom{n}{\omega}}.
\end{equation*}

Therefore, the probability of the whole algorithm succeeding in one cycle, i.e., the reciprocal of $P$ is
\begin{equation*}
  \frac{1}{P} = 1-(1-\varepsilon)^{|L_1|\times|L_2|\times|L_3|\times|L_4|} \approx \min(1,\varepsilon\times|L_1||L_2||L_3||L_4|).
\end{equation*}

In~\cite{EB22}, there are two specific methods for solving the nearest neighbor search problem, namely the Meet-In-The-Middle and the Indyk-Motwani method, which is related to the time complexity for constructing sets $L_{12},L_{34}$ and $L_{1234}$.
The analysis here is based on the Meet-In-The-Middle method, the other one is similar.
The time complexity described in Equation~(\ref{equ:time-DOOM})~can be written as
\begin{equation}\label{equ:time-DOOM2}
\begin{aligned}
  T &= P\times{T_G} + P\times\max\left(1,(R\cdot q)^{-1}\right)\left(T_{12}+T_{34}+T_{1234}\right),\text{ where }\\
  T_{12} &= \binom{l_1/2}{\omega_{11}/2}|L_1|+\binom{l_1/2}{\omega_{11}/2}|L_2|+\frac{|L_1||L_2|\binom{l_1}{\omega_{11}}}{2^{l_1}},\\
  T_{34} &= \binom{l_1/2}{\omega_{11}/2}|L_3|+\binom{l_1/2}{\omega_{11}/2}|L_4|+\frac{|L_3||L_4|\binom{l_1}{\omega_{11}}}{2^{l_1}},\\
  T_{1234} &= \binom{l_2/2}{\omega_{2}/2}|L_{12}|+\binom{l_2/2}{\omega_{2}/2}|L_{34}|+\frac{|L_{12}||L_{34}|\binom{l_2}{\omega_{2}}}{2^{l_2}},\text{ and }\\
  &|L_{12}| = \frac{|L_1||L_2|\binom{l_1}{\omega_{11}}}{2^{l_1}},\quad\quad|L_{34}| = \frac{|L_3||L_4|\binom{l_1}{\omega_{11}}}{2^{l_1}}.
\end{aligned}
\end{equation}

To minimize the time complexity, the base sets should be equal in size, i.e. $|L_1| =|L_2| =|L_3| =|L_4| = \sqrt[4]{\binom{k}{p}}$.

\subsection{Decoding one out of many for advanced ISD variants}

Now that we need to solve the syndrome decoding problem for a set of $S = \{\bs_i,1\leq i\leq N\}$ for $N$ independent instances.
Algorithm~\ref{alg:doom-frame} presents an algorithm for solving one out of $N$ problems for advanced variants, which is very similar to Algorithm~\ref{alg:gene-frame}.
But here the size of the sets $L_i$ depend on the parameter $N$.

\begin{algorithm}[h]
\caption{DOOM ISD Framework (Adapted from~\cite{sendrier11,EB22})}\label{alg:doom-frame}
\begin{algorithmic}[1]
    \Require parity check matrix $H\in\F_2^{(n-k)\times n}$, syndrome set $ S = \{\bs\}$, error weight $\omega\in[n]$
    \Ensure $\be\in\F_2^{n}$, which satisfies $H\be \in S$ and $wt(\be) = \omega$
    \Repeat
    \State choose random permutation matrix $P\in\F_2^{n\times n}$
    \State apply Gaussian elimination to transform $HP$ and all elements $\bs$ in $S$ into Equation~(\ref{mat:H-pre})
    \ForAll{$e_1\in L_1$ and $e_2\in L_2$}
    {
    \\
    \quad\quad\quad compute set $L_{12}$ which has the form
    \[
    L_{12} \leftarrow \{\be_1 + \be_2|(\be_1,\be_2)\in L_1\times L_2\text{ and }wt(\tilde{H}_1\be_{1}+\tilde{H}_1\be_{2}) = \omega_{11}\}
    \]
    }
    \EndFor
    \ForAll{$e_3\in L_3$ and $e_4\in L_4$}
    \ForAll{$\tilde{\bs} = (\tilde{\bs_1},\tilde{\bs_2},\tilde{\bs_3})\in S$}
    \\
    \quad\quad\quad\quad\quad compute set $L_{34}$ which has the form
    \[
    L_{34} \leftarrow \{\be_3 + \be_4|(\be_3,\be_4)\in L_3\times L_4\text{ and }wt(\tilde{H}_1\be_{3}+\tilde{H}_1\be_{4})+ \tilde{\bs_1} = \omega_{11}\}
    \]
    \EndFor
    \EndFor
    \ForAll{$\bx_1\in L_{12}$ and $\bx_2\in L_{34}$}
    {
    \\
    \quad\quad\quad compute set $L_{1234}$ which has the form
    \begin{equation*}
    L_{1234} \leftarrow \{\bx_1+\bx_2|(\bx_1,\bx_2)\in L_{12}\times L_{34}\text{ and }wt(\tilde{H}_2\bx_{1}+\tilde{H}_2\bx_{2})+ \tilde{\bs_2} = \omega_{2}\}
    \end{equation*}
    }
    \EndFor
    \For{$\bx\in L_{1234}$}
    \\
    \quad\quad\quad compute $(\bx,\tilde{H}\bx+\tilde{\bs})$
    \If{$wt(\bx,\tilde{H}\bx+\tilde{\bs}) = \omega$}
    \\
    \quad\quad\quad\quad\quad Break
    \EndIf
    \EndFor
    \Until{$wt(\bx,\tilde{H}\bx+\tilde{\bs}) = \omega$}
    \State The corresponding $P(\bx,\tilde{H}\bx+\tilde{\bs})$
\end{algorithmic}
\end{algorithm}

Assume that $P_N$ random permutations are needed to obtain the weight distribution.
In this case, the probability of the algorithm succeeding in one cycle is
\begin{equation*}
  \frac{1}{P_N} = 1-(1-\varepsilon)^{N\times|L_1|\times|L_2|\times|L_3|\times|L_4|} \approx \min(1,\varepsilon\times N|L_1||L_2||L_3||L_4|),
\end{equation*}
which is equal to $N\times \frac{1}{P}$ when $\varepsilon\times N|L_1||L_2||L_3||L_4|<1$.
We then examine in detail the time complexity.
Let $|L_1| = |L_2| = |L_3|$ and $|L_1|$ and $|L_4|$ is adjustable with $|L_1|^3|L_4| = \binom{k}{p}$.

For one out of $N$ syndrome decoding problem, the total time complexity is
\begin{equation*}
  T_N = P_N\times{(T_G)} + P_N\times\max\left(1,(R\cdot q)^{-1}\right)\left(T_{12}+T_{34}+T_{1234}\right),
\end{equation*}
In detail, the $T_{12}+T_{34}+T_{1234}$ section can be written as follows
\begin{equation*}
\begin{aligned}
  \binom{l_1/2}{\omega_{11}/2}|L_1|+\binom{l_1/2}{\omega_{11}/2}|L_2|+|L_{12}| +&\\
  \binom{l_1/2}{\omega_{11}/2}|L_3|+\binom{l_1/2}{\omega_{11}/2}N|L_4|+|L_{34}| +&\\
  \binom{l_2/2}{\omega_{2}/2}|L_{12}|+\binom{l_2/2}{\omega_{2}/2}|L_{34}|+\frac{|L_{12}||L_{34}|\binom{l_2}{\omega_{2}}}{2^{l_2}}&,\\ \text{ where }
  |L_{12}| = \frac{|L_1||L_2|\binom{l_1}{\omega_{11}}}{2^{l_1}},\quad\quad |L_{34}| =& \frac{N|L_3||L_4|\binom{l_1}{\omega_{11}}}{2^{l_1}}.
\end{aligned}
\end{equation*}
Then, it can be transformed into
\begin{equation*}
  \begin{aligned}
  \binom{l_1/2}{\omega_{11}/2}\left(3|L_1|+N\frac{\binom{k}{p}}{|L_1|^3}\right) &+ \frac{\binom{l_2/2}{\omega_2/w}\binom{l_1}{\omega_{11}}}{2^{l_1}}\left(|L_1|^2+\frac{N\binom{k}{p}}{|L_1|^2}\right) \\
  & + \frac{N\binom{k}{p}\binom{l_1}{\omega_1}^2\binom{l_2}{\omega_2}}{2^{l_2-2l_1}}.
  \end{aligned}
\end{equation*}

At the end, $T_{12}+T_{34}+T_{1234}$ can be minimized by making $|L_1| = \sqrt[4]{N\binom{k}{p}}$, and $|L_4| = \frac{\sqrt[4]{\binom{k}{p}}}{\sqrt[3/4]{N}}$.
Thus, we deduce $T_N \leq \frac{T}{\sqrt{N}}$.
The result is similar to that of Sendrier~\cite{sendrier11} and confirms the folklore conjecture.
That is, considering $N$ independent instances, the time complexity to solve one out of $N$ instances using these advanced ISD algorithm is approximately $1/\sqrt{N}$ times the time complexity to solve a single instance.

\section{Conclusion}

In this work, we explore decoding problem with multiple instances in two different scenarios.
In the first scenario, we employ a direct preprocessing strategy that divides the current binary search tree framework into a preprocessing phase and an online computation step. The modification reduce the amortized time complexity of solving each of the $N$ instances, and affects the concrete security of some code based post-quantum schemes. Finding a better way to employ the preprocessing strategy is an interesting problem for further research.
In the second scenario, we consider the decoding one out of many problem. We combine Sendrier's and Esser-Bellini's framework to examine the time complexity of solving one out of the $N$ instances. The result confirms the folklore conjecture that
the advanced variants can be sped up by a factor of roughly $\sqrt{N}$.

\section*{Acknowledgements}
This work was partially supported by the National Key Research and Development Program of China(Grant No. 2021YFB3100200) and the Major Basic Research Project of Natural Science Foundation of Shandong Province, China (Grant No. ZR202010220025).

\bibliographystyle{splncs03}
\bibliography{dec}
\end{document}